\documentclass[a4paper,11pt]{article}
\usepackage{jinstpub} 
\usepackage{lineno}

\title{Control and read-out of the HEPD-02 tracking system onboard CSES-02 satellite}

\author[a]{S.~Bartocci,}
\author[b,c]{R.~Battiston,} 
\author[d,e]{S.~Beolè,} 
\author[e]{F.~Benotto,} 
\author[f]{P.~Cipollone,} 
\author[e]{S.~Coli,} 
\author[g,h]{A.~Contin,} 
\author[i]{M.~Cristoforetti,} 
\author[f]{C.~De Donato,} 
\author[f]{C.~De Santis,} 
\author[i]{A.~Di Luca,} 
\author[e]{F.~Dumitrache,} 
\author[b,c]{F.~M.~Follega,} 
\author[f]{S.~Garrafa Botta,} 
\author[b,c]{G.~Gebbia,} 
\author[b,c]{R.~Iuppa,} 
\author[i]{A.~Lega,} 
\author[h]{M.~Lolli,} 
\author[f]{G.~Masciantonio,} 
\author[j]{M.~Mergè,} 
\author[k,m]{M.~Mese,} 
\author[b,c]{R.~Nicolaidis,} 
\author[c]{F.~Nozzoli,} 
\author[h]{A.~Oliva,} 
\author[m]{G.~Osteria,} 
\author[f]{F.~Palma,} 
\author[g,h]{F.~Palmonari,} 
\author[k,m]{B.~Panico,} 
\author[d,e]{S.~Perciballi,} 
\author[m]{F.~Perfetto,} 
\author[n,f]{P.~Picozza,} 
\author[h]{M.~Pozzato,} 
\author[b,c]{E.~Ricci,} 
\author[o]{M.~Ricci,} 
\author[p]{S.~B.~Ricciarini,} 
\author[g,h]{Z.~Sahnoun,} 
\author[d,e]{U.~Savino,} 
\author[k,m]{V.~Scotti,} 
\author[c]{E.~Serra,} 
\author[f]{A.~Sotgiu,} 
\author[n,f]{R.~Sparvoli,} 
\author[q]{P.~Ubertini,} 
\author[b,c]{V.~Vilona,} 
\author[j]{S.~Zoffoli}
\author[b,c,1]{and P.~Zuccon\note{Corresponding author.} }

\affiliation[a]{INFN - AC, V. E. Fermi 54, 00044 Frascati (RM), Italy}
\affiliation[b]{Università di Trento,  V. Sommarive 14, 38123, Trento, Italy}
\affiliation[c]{INFN - TIFPA, V. Sommarive 14, 38123, Trento, Italy}
\affiliation[d]{Università di Torino, Via P.Giuria 1, 10125, Turin, Italy}
\affiliation[e]{INFN - Sezione di Torino, Via P.Giuria 1, 10125, Turin, Italy}
\affiliation[f]{INFN - Sezione di Roma Tor Vergata, V. della Ricerca Scientifica 1, 00133, Rome, Italy}
\affiliation[g]{Università di Bologna, V.le Berti Pichat 6/2, Bologna, Italy}
\affiliation[h]{INFN - Sezione di Bologna, V.le Berti Pichat 6/2, Bologna, Italy}
\affiliation[i]{Fondazione Bruno Kessler, Via Sommarive 18, 38123, Trento, Italy}
\affiliation[j]{Italian Space Agency, V. del Politecnico, I-00133 Rome, Italy}
\affiliation[k]{Università degli Studi di Napoli Federico II, V. Cintia, 80126, Naples, Italy}
\affiliation[m]{INFN - Sezione di Napoli, V. Cintia, 80126, Naples, Italy}
\affiliation[n]{Università di Roma Tor Vergata, V. della Ricerca Scientifica 1, 00133, Rome, Italy}
\affiliation[o]{INFN - LNF, V. E. Fermi 54, 00044 Frascati (RM), Italy}
\affiliation[p]{IFAC-CNR, Via Madonna del Piano 10, 50019 Sesto Fiorentino (FI), Italy}
\affiliation[q]{INAF-IAPS, V. Fosso del Cavaliere 100, 00133, Rome, Italy}

\emailAdd{paolo.zuccon@unitn.it}

\abstract{The High Energy Particle Detector (HEPD-02) is a payload of the second China Seismo-Electromagnetic Satellite (CSES-02), designed and built by the Italian Limadou collaboration. Its purpose is to detect cosmic rays and trapped particles of radiation belts, in the kinetic energy range 3-100 MeV for electrons, 30-200 MeV for protons. HEPD-02 is the first space detector to use a tracking detector based on  Monolithic Active Pixel Sensors (MAPS). The MAPS provides high spatial resolution, low noise, increased robustness, and low production costs. Operating MAPS in space presents a significant challenge due to strict power consumption requirements. To meet such constraints, a custom Tracker Data Acquisition (TDAQ) board and firmware have been designed and implemented, by using a commercial low-power Field Programmable Gate Array (FPGA). This paper addresses the design features of the TDAQ unit, enabling the tracking detector to be operated efficiently, with particular focus on the power consumption performance.}

\begin{document}
\maketitle
\flushbottom

\section{Introduction}


The China Seismo-Electromagnetic Satellite CSES-02 is part of a scientific multi-spacecraft program devoted to studying the electromagnetic, plasma and particle environment in the Low Earth Orbit (LEO) region, in a collaboration framework led by the China National Space Administration (CNSA) and the Italian Space Agency (ASI).

The most relevant objective of the CSES program is the observation of perturbations, involving the inner Van Allen belt and the Earth's atmosphere, originated by solar \cite{martucci2023first,palma2021high,palma2021august} or terrestrial phenomena, producing changes in the electromagnetic fields and in the population of trapped charged particles. CSES aims to establish possible statistical correlation between such transients and seismic phenomena\cite{aleksandrin2003high,sgrigna2005correlations}. The first satellite, CSES-01, was launched in 2018, whereas the second, CSES-02, has been launched on June 14th 2025.


The  High Energy Particle Detector HEPD-02 \cite{DeSantis:2021WO, Scotti:2019/9}, built by the Italian \textit{Limadou} collaboration, is one of the CSES-02 payloads, specifically designed to study cosmic-ray electrons and protons in the kinetic energy range from 3 to 100 MeV and from 30 to 200 MeV, respectively.
HEPD-02 is an upgraded version of the predecessor HEPD-01\cite{picozza2019hepd,ambrosi2018hepd} onboard CSES-01. It is capable of performing event-based particle identification through a simultaneous measurement of the incoming particle direction and energy.



\begin{figure*}[t]
    \centering
    \includegraphics[width = 1. \linewidth]{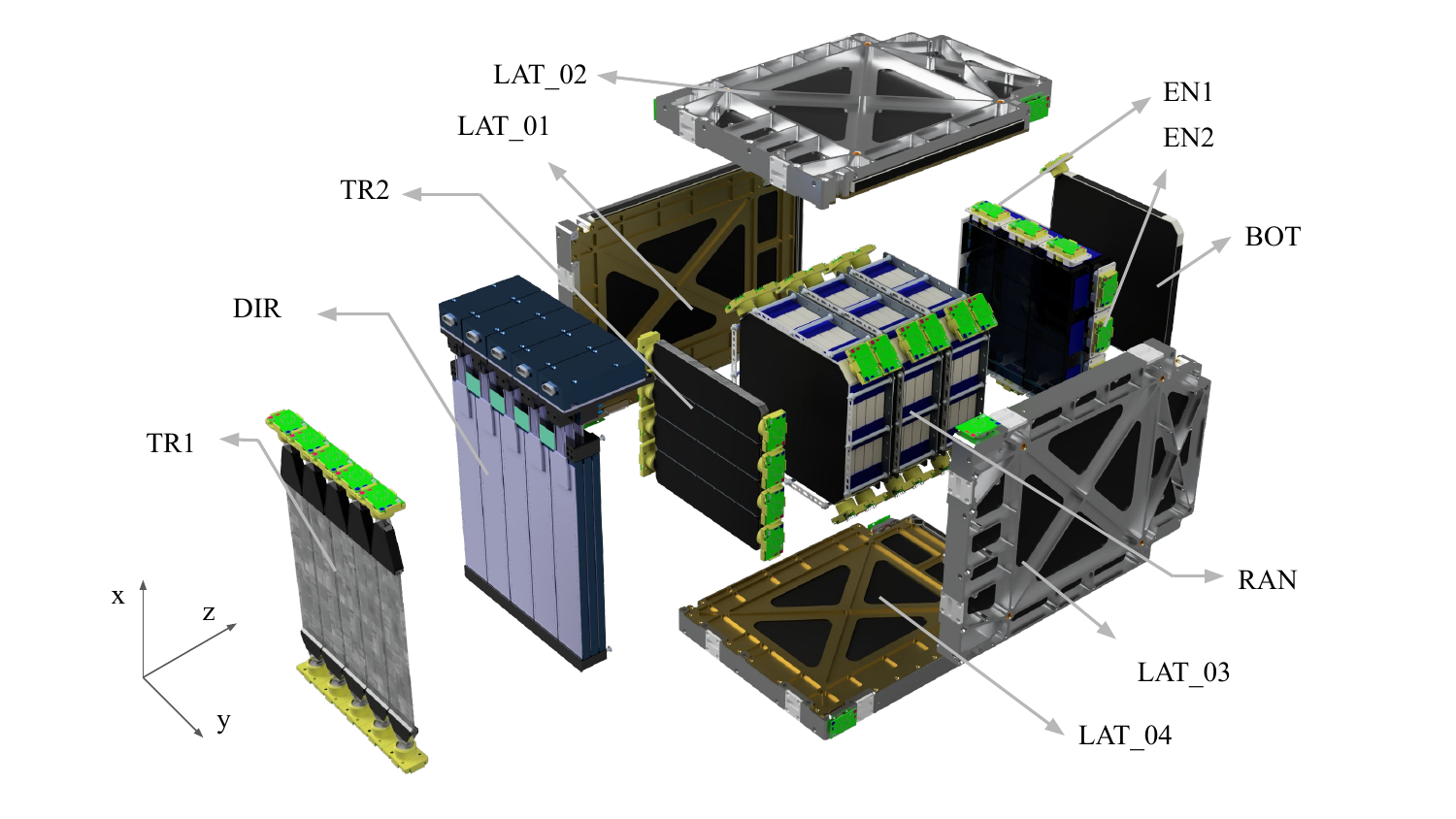}
    \caption{Exploded view of the HEPD-02 detectors: trigger planes (TR1 and TR2), silicon pixel tracker (DIR), range calorimeter (RAN), energy calorimeter (EN1, EN2), bottom (BOT) and lateral (LAT\_01..04) containment detectors.}
    \label{fig:HEPD_assembly}
\end{figure*}

HEPD-02 is composed of a set of specialized particle detectors, \textcolor{black}{depicted in Fig.~\ref{fig:HEPD_assembly}, which are installed in a volume of dimensions 37 cm $\times$ 53 cm $\times$ 39 cm together with the relevant electronics.} At the top side, facing the zenith direction during orbital flight, the tracker - or direction detector (DIR) -\cite{iuppa2021innovative, ricciarini2021enabling, coli2021sissa} is composed of three layers of Monolithic Active Pixel Sensors (MAPS) described in Sect.~\ref{sect:Tracker}. Above and below the DIR, the trigger planes TR1 and the TR2 are composed of mutually orthogonal bars of plastic scintillators; by design, the geometrical dimensions of the TR1 five bars match the five underlying DIR detector elements.


This front part is followed by the range calorimeter (RAN) made of 12 square plastic scintillator elements and by the energy calorimeter (EN) made of two layers of LYSO (lutetium-yttrium oxyorthosilicate) crystals. Scintillator panels at the bottom (BOT) and on the sides (LAT\_01 to LAT\_04) surround the calorimeter with the main purpose of tagging events with particles not fully contained. The TR1, TR2, RAN, EN1, EN2, BOT and LAT units are read-out by photomultiplier tubes (PMTs), for a grand total of 64 PMTs.


The detector control and data acquisition (DAQ) is implemented in three distinct electronic units.

The data processing and control unit (DPCU)\cite{masciantonio2021hepd} is the main HEPD-02 computer, controlling data interfaces with satellite and configuring the entire instrument. In particular, the DPCU collects the scientific data produced by the detectors and arranges them for transmission to the satellite via a dedicated RS-422 link.

The trigger (TRIG) board \cite{instruments7040053, scotti2023trigger} is dedicated to the read-out of PMT signals and to initiate the general event DAQ when an incoming particle activates specific patterns of hit PMTs.

The tracker data acquisition (TDAQ) board is dedicated to the read-out of DIR. In Sect.~\ref{TDAQ} and  following ones, the design of the TDAQ board and the solutions implemented in its functional architecture are discussed, with particular attention to the strategies for keeping the combined DIR and TDAQ power consumption within the limited budget allowed by the specific satellite application.





\section{The silicon pixel tracker}
\label{sect:Tracker}
The HEPD-02 direction detector (DIR)  is the first silicon pixel tracker ever deployed in space, marking a major advancement in particle spectrometry for space applications. Unlike the microstrip detectors used in previous missions such as AMS-02\cite{ALPAT2010207} and PAMELA\cite{STRAULINO2004168}, HEPD-02 employs Monolithic Active Pixel Sensors (MAPS), which offer key benefits including higher spatial resolution, reduced noise, lower power consumption, a more compact form factor, and cost-effective fabrication. \textcolor{black}{Further details on the design, construction and qualification of the DIR detector can be found in \cite{hepd02_dir}.}

HEPD-02 DIR uses the ALTAI MAPS sensor, selected for its excellent noise and resolution performance, low power requirements, and ultra-thin  substrate, which minimizes multiple scattering. The choice also builds on the extensive experience and documentation from the ALICE experiment at CERN\cite{ALPIDE_structure}, which uses the closely related ALPIDE sensor. \textcolor{black}{One of the most relevant differences is the case use, several simultaneous tracks at trigger rate of $\sim$ 40 MHz for ALICE and typically single or few tracks at few kHz for HEPD-02.}



The ALTAI sensor is fabricated by Tower Semiconductor LTD with a 180 nm CMOS process over a silicon substrate of 50 $\mu$m only. It features an array of 512 $\times$ 1024 pixels over an area of 15 $\times$ 30 mm$^2$, with $\sim$28 $\mu$m  pitch in both directions, thus providing a very high spatial resolution. Each pixel cell contains a sensing diode read-out by an analog circuit (amplifier, shaper and threshold discriminator) that generates a binary hit information. This is then stored and processed in the digital part of the device.



\begin{figure*}[h]
    \centering
    \includegraphics[width = 1.\linewidth]{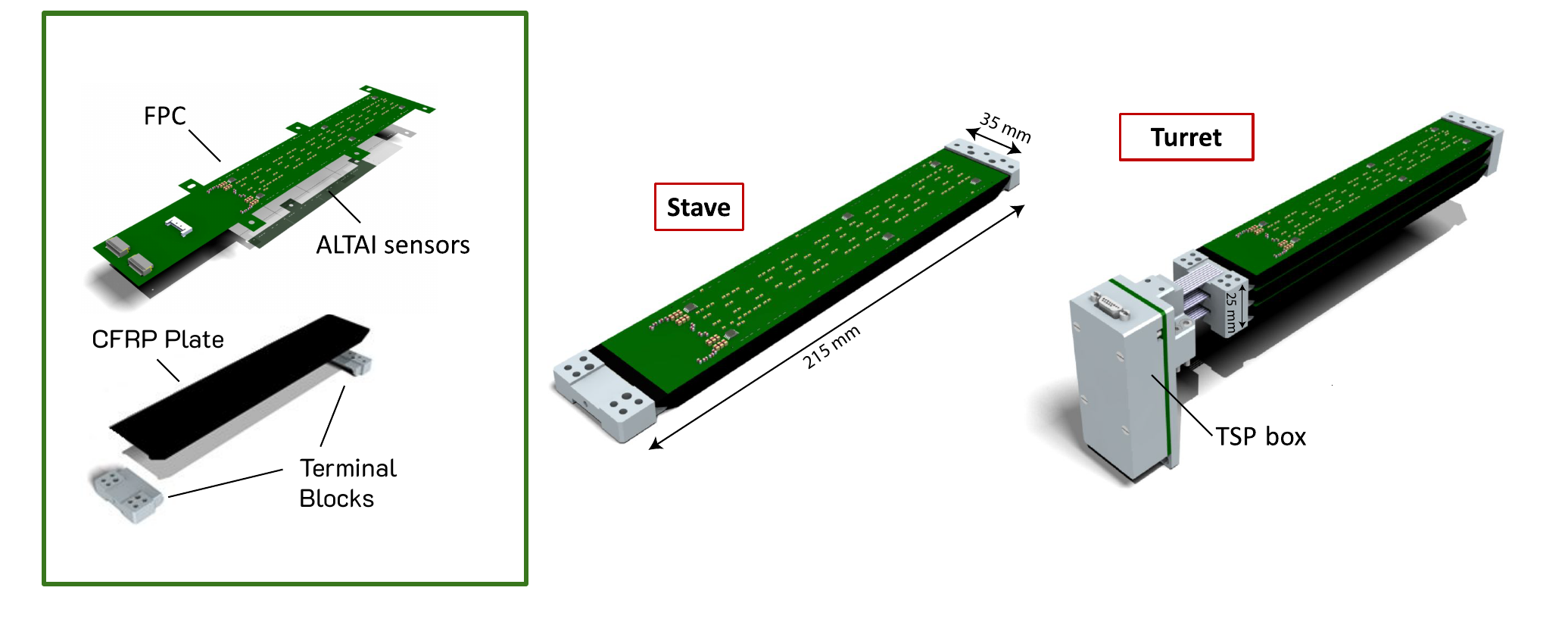}
    \caption{Visualization of DIR turret assembly. In a turret, 3 staves are vertically stacked and connected to the lateral TSP board. Each stave contains 2 columns of 5 ALTAI chips. }
 
    \label{fig:turret}
\end{figure*}

The DIR is composed of five identical modules called \textit{turrets}, arranged side by side, covering an area of 150 $\times$ 150 mm$^2$. Each turret (see Fig.~\ref{fig:turret}) is  an independent direction detector, made of a stack of three MAPS layers, vertically spaced by 8.5 mm. A small lateral tracker splitter (TSP) board, equipped with Glenair Micro-D connectors, constitutes the interface with the HEPD-02 power supply unit, delivering circuit power and silicon bias, and with the TDAQ board for digital lines.


The turret stack element, called \textit{stave}, is the fundamental mechanical and electrical unit of the DIR. The stave is formed by a custom-designed U-shaped carbon-fiber reinforced polymer (CFRP) plate \cite{coli2021sissa} supporting 10 ALTAI chips, organized in two columns of 5 chips each, for a total area 30 x 150 mm$^2$. The CFRP plate provides structural strength, capable of withstanding the stresses expected during launch and orbital maneuvers; it also offers good thermal conductivity for transportation of the heat generated by the MAPS towards the aluminum terminal blocks, which constitute the mechanical and thermal interface of the stave with HEPD-02 structure.

A flexible printed circuit (FPC) is glued on top of the ALTAI chips, providing the routing of power, bias and data lines. Ultrasonic wire bonding is used to obtain the electrical connections between the FPC and the underlying ALTAI bonding pads, that are accessible through corresponding FPC holes. For redundancy, each connection is performed with three bonding wires.


For each column, one ALTAI chip is configured as master and the other 4 chips as slaves. Each slave is connected with the master by means of a multi-drop LVDS (MLVDS) input clock, a CMOS bidirectional control data line and 4 CMOS read-out lines operating in double data rate mode, thus providing 8 bit per clock cycle. The master is in turn connected with the TDAQ via 2 MLVDS lines (input clock and bidirectional serial data); \textcolor{black}{the two master ALTAI on a stave share the same physical connections with TDAQ, which means that the serial data line is actually a shared bus.}


\section{The Tracker Data Acquisition (TDAQ) board}
\label{TDAQ}

\begin{figure}[htb]
    \centering
    \includegraphics[width = 1 \linewidth]{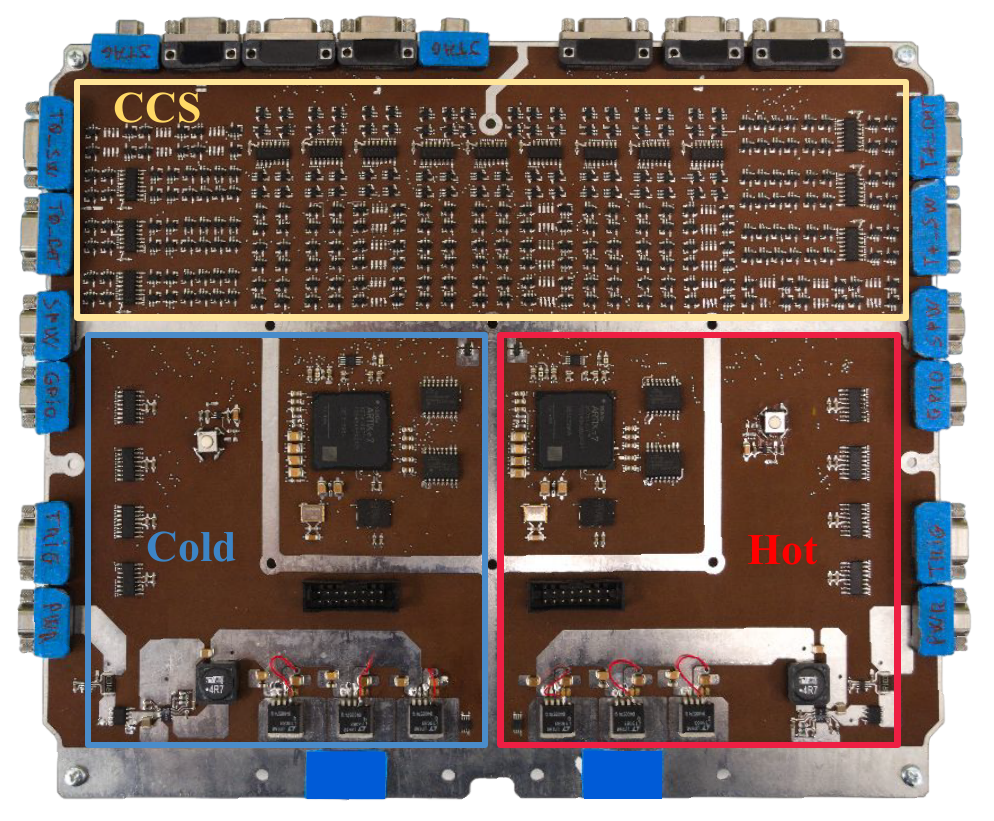}
    \caption{Visualization of the TDAQ board. In the illustration, the hot/cold redundancy is emphasized by the red/blue contours. }
    \label{fig:TDAQ_board}
\end{figure}
\begin{figure*}[htb]
    \centering
    \includegraphics[width = 0.9 \linewidth]{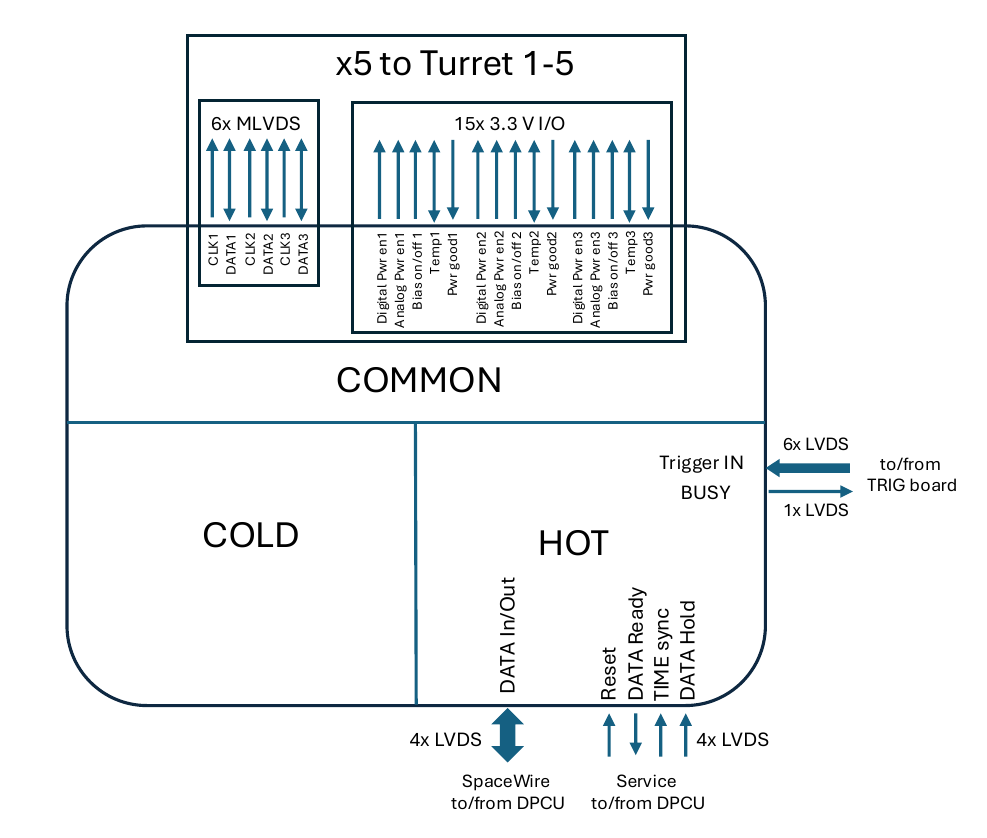}
    \caption{Schematic of the TDAQ connections with the DIR detector, the DPCU, and the TRIG boards. The connections on the hot section are also present in the cold section (not drawn). }
    \label{fig:TDAQ_conn}
\end{figure*}
\begin{figure*}[htb]
    \centering
    \includegraphics[width = 1. \linewidth]{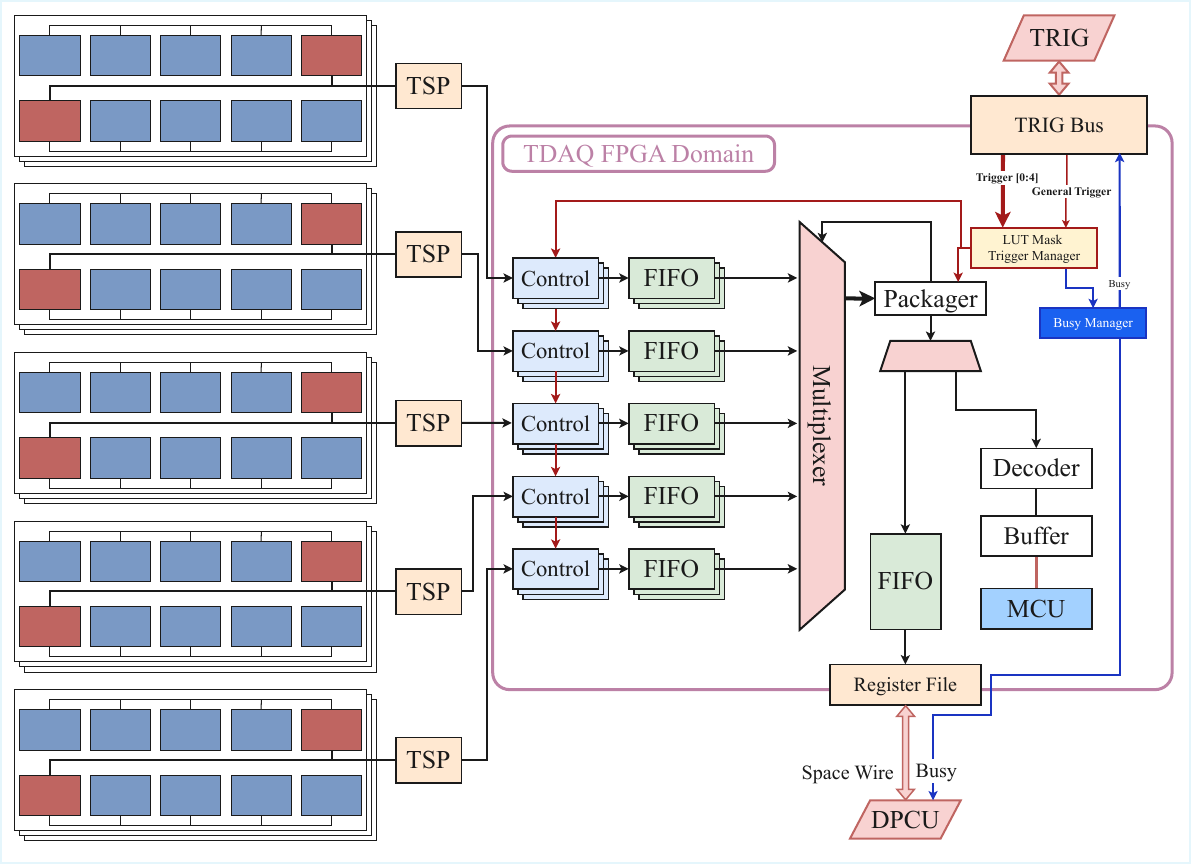}
    \caption{Schematical view of the DIR with the connections to the TDAQ board, ALTAI Master (red squares) chips,  ALTAI slaves (blue squares). The internal architecture of the TDAQ firmware is depicted on the right. }
    \label{fig:TDAQ_firm}
\end{figure*}


The TDAQ board has been implemented with a custom design of both hardware and firmware, with a parallel architecture for a fast enough read-out speed, while keeping the board power consumption within the required limits. The preliminary design of TDAQ was presented in a short conference report \cite{Nicolaidis:2023l9}, while in this paper we discuss the motivations for a custom design and the details of its implementation.

The primary task of the TDAQ board is to operate the DIR detector by acquiring pixel data, delivering them to the DPCU and managing the ALTAI configuration and calibration tasks. The TDAQ board is also responsible for controlling the on/off state of power and bias for each of the 15 DIR staves, by operating solid-state switches located on the TSPs.




The TDAQ board circuitry is composed of three parts: the  Common Connections Section (CCS), the Hot Digital Section (HDS) and the Cold Digital Section (CDS), as depicted in Fig.~\ref{fig:TDAQ_board}. \textcolor{black}{The HDS and the CDS are identical, according to a hot/cold redundancy scheme, like for most of the HEPD-02 electronics: the HDS is normally operating, while the CDS is normally kept powered-off. In case of unrecoverable failure of HDS, the DPCU will switch HEPD-02 to safe mode and wait for a command issued from ground to change the payload configuration, enabling the CDS instead of the HDS.}

The CCS hosts the interface with DIR, with two connections (data and service) for each turret, implemented with Glenair Micro-D GMR7590 connectors (fig. \ref{fig:TDAQ_conn}).
The data connection carries six MLVDS lines, implemented with Texas Instruments SN65LVDM051 high-speed transceivers. Each stave of the turret uses two lines: one for the clock output and one for bidirectional serial data.
The service connector hosts single digital signals to manage the stave: analog-part power enable, digital-part power enable, power good status, silicon bias on/off, and a 1-wire line for temperature sensor chain read-out.
For each line, a specially designed circuit, built with two complementary switches (Analog Device ADG701), automatically handles the correct connection to either HDS or CDS, according to which one is currently powered.

The HDS/CDS contains the board control logics, the communication interfaces with DPCU and TRIG boards (both hot/cold redundant) and the internal power distribution network.
The control logics is implemented on a FPGA from the Xilinx Artix 7 series, characterized by low power consumption (L1 type, 0.95 V core voltage) and wide operating range for junction temperature (-40$^{\circ}$C to +100$^{\circ}$C).
Two 16 MB flash memories, Cypress S25FL256S, are present: the first stores the FPGA configuration code, and the second acts as a mass storage for the soft MicroBlaze processor instantiated within the FPGA (described in Sect.~\ref{Firmware}). An auxiliary 4 MB SRAM (Cypress CY7C1071DV33) is also present.
The connections with the DPCU and TRIG boards (fig. \ref{fig:TDAQ_conn}) are implemented using standard LVDS channels, driven by Texas Instruments transceivers of the SN65LVDS3x family, \textcolor{black}{for discrete control signals and data link, which are described in Sect. \ref{Firmware}.}


The selected components comply with the automotive or industrial qualification and were successfully used for several years during previous low-Earth space missions, such as HEPD-01. In particular, the Xilinx Artix 7 FPGA features adequately small cross sections for SEU (Single Event Upset) on configuration memory and programmable logic bits \cite{tambara} as well as stable operation with no functional errors in all critical parts irradiated with doses up to 2 Mrad \cite{lentaris}. \textcolor{black}{Given the irradiation conditions for the CSES-02 mission, even taking into account enhancements driven by solar activity, the expected SEU rate for the whole TDAQ amounts to less than 1 event in 10 years.}

\section{The TDAQ firmware}
\label{Firmware}

The TDAQ firmware \textcolor{black}{is implemented in the Xilinx Artix 7 FPGA and} includes a finite state machine (FSM) section, for data acquisition and external interfaces with DPCU and TRIG boards, and a soft microcontroller unit (MCU) for executing the more complex calibration tasks of ALTAI chips. The structure of the TDAQ firmware is shown in Fig. \ref{fig:TDAQ_firm}.


The FSM section has been fully developed in VHDL (Standard IEEE 1076-2019) and profits of the reliability of circuits implemented on the FPGA fabric. The design is based on modularity and parallelization. During the development, the design modularity allowed to readily adapt to changes in the detector requirements and ensured high reusability across different devices within the same family. Parallelization, a notable strength of FPGA-based firmware, has been leveraged to address the relatively slow communication speed with the ALTAI sensors, as discussed later. The synthesis, implementation, testing, and installation of the firmware were all carried out using the Vivado suite provided by Xilinx.


The TDAQ is designed to operate either in DAQ or idle mode, according to the general operation of HEPD-02, which dedicates most of its time to the acquisition of particle events, with interspersed phases for instrument calibration or self-testing.

In DAQ mode, the TDAQ responds to trigger signals from the TRIG board, i.e.~the general trigger and 5 additional signals indicating the single activated TR1 bars; it sends back a busy signal, inhibiting further triggering until the DIR data (i.e.~the information on hit pixels) have been fully collected and processed.

Stave data are read-out in parallel: for each of the 15 staves, the corresponding Control module sends a read-out command to the master ALTAI chips; after a suitably tuned time delay, it collects the data and puts them in an internal FIFO buffer. \textcolor{black}{The time interval between trigger pulse and data collection is thus fixed (with a negligible synchronization jitter) and chosen in such a way to match the time window for which the ALTAI threshold discriminator binary output is kept to 1 after each hit.}

The Packager module acts as event builder, asynchronously pulling data from the 15 Control modules and rearranging them into a DIR event packet, optimized for space saving and with a Cyclic Redundancy Check (CRC) word for data consistency check. \textcolor{black}{The DIR event packet includes an event number and time stamp of trigger reception, to ensure unique correlation with the corresponding TRIG event packet.} The DIR event packets are stored in a FIFO output buffer for subsequent asynchronous read-out by DPCU\textcolor{black}{, while the busy signal is negated, except in case the buffer is full because of the accumulation of event packets to be read.} \textcolor{black}{The DIR event packets have variable size, of the order of few hundreds bytes, depending on the pattern of hit pixels.} \textcolor{black}{The time needed after trigger reception, for data read-out, processing and storage on FIFO output, increases with the number of hit pixels; for typical events with single or few tracks it amounts to few hundreds $\mu$s, fully compatible with the trigger rates of few kHz maximum which can be managed by HEPD-02.}


 
In DAQ mode the TDAQ responds only a reduced set of commands from DPCU, such as reading event packets or the essential diagnostic information to assess the correct DIR operation.

In idle mode, the TDAQ serves all commands from DPCU, including read/write operations on any register, configuration operations and requests for different kinds of ALTAI calibrations or self-test operations; \textcolor{black}{these last more complex commands are executed autonomously by the MCU, which operates on the data diverted from the Packager to a Decoder module and then to a MCU accessible RAM buffer.}

%
%
%
%

The interface with DPCU is established by exposing a shared memory segment, called register file, that can be read or written by the DCPU. In particular, the register file contains a command register, where the DPCU can write a code corresponding to the action to be executed by the TDAQ; the command output is then available in a dedicated section of the memory segment. Two other register file notable sections contain the first available event packet and the status/health data of DIR and TDAQ.

DPCU and TDAQ communicate using the SpaceWire Lite protocol (IEEE 1355-1995)\cite{spacewire}, implemented over a full-duplex SpaceWire standard digital connection \textcolor{black}{driven at 20 Mbps}, where DPCU acts as master and TDAQ as slave. Additional single digital lines are used as TDAQ service inputs (board reset, time counter synchronization, event packaging hold) and output (event packet ready).


%

The MCU has been implemented using a Microblaze soft micro-controller. Its purpose is to execute tasks involving data processing to determine the optimal configuration parameters for each of the 150 ALTAI chips and to execute calibrations. The ability to write standard code for these tasks and use flash memory for mass storage broadens the scope of operations that can be conducted in-flight, offering enhanced flexibility to adapt to unforeseen post-launch conditions.
With respect to a multi-core ARM processor, this solution has the advantage of configuring the features according to the needed set of tasks; furthermore, when not in use, the soft microcontroller can be disabled, minimizing the power consumption.

The MCU is complemented by other Xilinx IPs, including two AXI Timers, the AXI External Memory Controller, and the AXI Quad SPI for interface with the flash memories. 
To enhance system reliability, an AXI Timebase Watchdog Timer was integrated into the design together with the MCU. Its purpose is to prevent the software running on the MCU from freezing: if the MCU does not regularly reset the watchdog timer, a signal is generated that triggers a soft reset of the MCU.


The TDAQ firmware adopts mitigation techniques for single-event upsets (SEU) which may be induced by radiation in the LEO environment. For the FPGA internal SRAM configuration memory, we implemented the Soft Error Mitigation Controller (SEM) IP core by Xilinx\cite{semxilinx, bates2018single}. The SEM continuously reads the configuration memory and performs integrity checks. It can thus detect errors and automatically rewrite the affected configuration data.

Even after the configuration memory is restored, the FSM may be stuck in an invalid state, thus preventing proper device functioning.
To handle such situations, the TDAQ FSM can set specific error flags in the status/health data regularly monitored by the DPCU: in such cases, the DPCU may apply a reset of the logic or eventually reconfigure the FPGA.



Additional mitigation techniques for SEU consequences are adopted for FSM state registers, which are configured with Hamming(7,4) encoding, i.e.~with automatic correction of single bit upset.
The MCU core memory and the FIFO/RAM buffers use a similar Error Correcting Code to automatically fix single bit upsets.


\section{Strategies For An Efficient Power Management}
\label{sect:PowerManagement}

The ALTAI chip is the non dual-use version of the ALPIDE chip, originally developed for the HL-LHC upgrade of the ALICE experiment, operated at CERN LHC collider. These devices are designed for fast data delivery but not optimized for power consumption. In fact, when used at CERN, the chips are cooled with a water circuit so as to draw as much power as needed. On the other hand, using them in space missions requires careful considerations about power consumption.

To deploy the ALTAI chips in HEPD-02, we developed a novel approach with respect to the ALICE experiment, aiming at reducing the power absorbed by the digital part while leaving the analog part unchanged, thus maintaining the excellent detection performances of ALTAI sensors and, at the same time, obtaining a combined power consumption of DIR and TDAQ compliant with the stringent 13 W budget allowed by the specific application in the CSES-02 satellite.



As a main design guideline, we exploited the fact that in a satellite experiment such as HEPD-02, the trigger rates and amount of information to be read-out (i.e.~hit pixels) are orders of magnitude smaller than in the ALICE experiment. In particular, HEPD-02 is configured to manage trigger frequencies up to few kHz at peak, with few hundreds of bytes of information produced by the whole tracker. This allowed to slow down the DIR read-out while still maintaining it compatible with the instrument operation.

We chose to operate ALTAI in the master/slave mode: each stave hosts 10 chips organized in two groups, each with a master and four slaves. The master is the sole interface with TDAQ board and the data transmission is serialized through it, with significant power saving and  reduction of the physical connections towards TDAQ. This is important given the strict geometrical constraints for wire routing in the compact HEPD-02 geometry. \textcolor{black}{Measurements of power consumption indicate that} the use of master/slave mode allows to reduce the overall DIR power consumption by $\sim$30\% \textcolor{black}{with respect to a solution employing stand-alone ALTAI chips, i.e.~from 22 W} to 15 W (1 W per stave).


The ALTAI device is designed for a serial read-out of master chips by means of the built-in fast (1200 Mbps / 400 Mbps) Serial Data Transmission Module (SDTM), accounting for a power consumption of $\sim$90 mW. We found that it is possible to implement an alternative read-out using the much slower Control Logic Bus (CLB), a half-duplex serial line with a bandwidth of 40 Mbps. At the same time, the ALTAI modular design allows to \textcolor{black}{keep the SDTM logics permanently powered off}.
The CLB is normally used to configure the chip and deliver the trigger, but it also gives access to the ALTAI output buffers and can be used as an alternative data read-out path. Since, in its original purpose, the CLB was not intended for sustained data reading, we had to design from scratch a new implementation of the ALTAI communication protocol over the CLB in the TDAQ FSM section. The time needed to transmit the event data to TDAQ is typically of a few hundreds $\mu$s, fully compatible with trigger rates.
With this solution, the overall DIR power consumption \textcolor{black}{is expected to be further reduced by $\sim$20\%, i.e.~from 15 W} to 12 W (0.8 W per stave).



  

As an additional measure to comply with the power limit, we introduced a time gating of the digital clock signal. In DAQ mode, the clock is normally kept off and turned on when the TDAQ receives a trigger signal; it is then kept running while the TDAQ sends a read-out command and the event data are read by the TDAQ, then it is turned off again. With clock gating, the absorbed power is high only for a very limited time after each trigger; on the other hand, when the clock is off the ALTAI digital part is not active and the overall DIR power consumption is expected to be reduced to $\sim$6 W (0.4 W per stave).
This measure is feasible because the ALTAI analog part remains always active: when a pixel collects charge, a signal with a shaping time of few $\mu$s is immediately formed, and a corresponding pixel-hit line is asserted for all the time this signal is above a given threshold. An incoming read-out command initiates the check of all pixel-hit lines, with the addresses of hit pixels added to the output data stream. On the other hand, the time passing from the particle crossing to the transmission of read-out command from TDAQ to a stave is of few hundreds  ns, still compatible with the shaping time.

For what said above, with clock gating the power consumption increases with the trigger frequency and duration of the clock gate itself. Laboratory measurements were performed (Sect.~\ref{sect:performances}) to accurately assess the trigger rate dependence of the absorbed power. The combined DIR and TDAQ power turned out to be well within the allowed 13 W budget.



If necessary, further power reduction can be achieved by working on the spatial domain. As already reported, each DIR turret is matched with a scintillation bar of the first trigger plane TR1, with the TRIG board generating 5 specialized trigger signals corresponding to the pattern of hit TR1 bars. The TDAQ firmware was configured in such a way that it is possible to enable a reduced-power operation, with read-out of just one turret or up to 3 turrets, centered on the hit TR1 bar, which is feasible for the majority of events, with only one TR1 hit.

\begin{figure*}[!ht]
    \centering
    \includegraphics[width = 0.8 \linewidth]{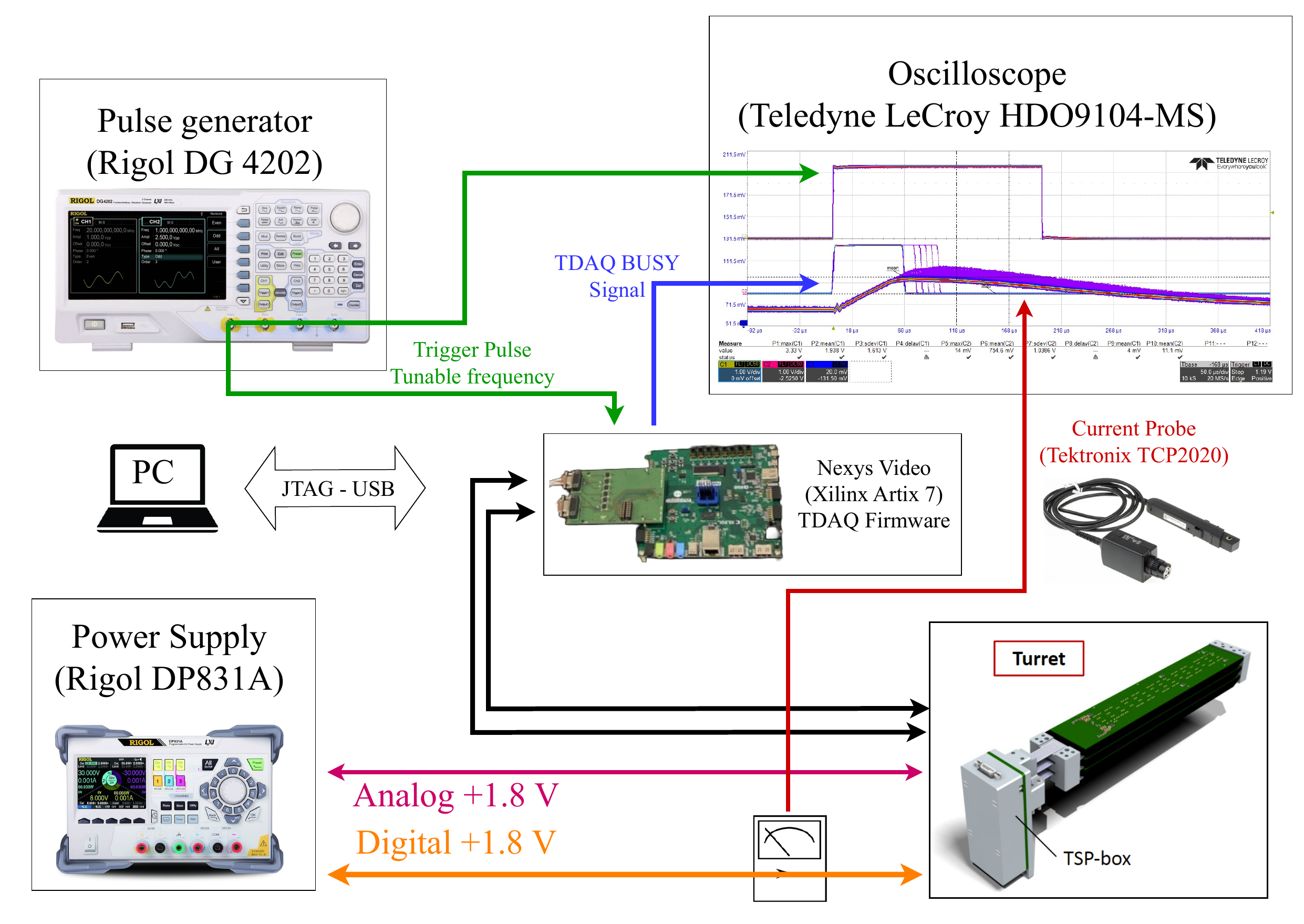}
    \caption{Schematic representation of the experimental setup recreated in the TIFPA laboratory for the characterization of the DIR power consumption. The TDAQ firmware is instantiated on the Xilinx Artix 7 FPGA mounted on the  Digilent's Nexys Video evaluation board. The connections between the FPGA board and the DIR turret (via the TSP) are made through  a custom expansion board. To study the power consumption versus the trigger frequency, the DAQ chain was triggered by a pulse generator. The average power consumption was measured \textcolor{black}{by the power supply unit (see text for details).}}
    \label{fig:PowerCons_Characterization}
\end{figure*}
 
\section{Measured power consumption performances}
\label{sect:performances}

We implemented an experimental setup (see Fig.~\ref{fig:PowerCons_Characterization}) to accurately measure the power needed to operate the DIR and thus confirm the effectiveness of the design features introduced to maintain the power consumption within the allowed budget.

A DIR turret was connected to a bench power supply unit (Rigol DP831A) providing the 1.8 V analog and digital power supply lines and measuring the current delivered on each output. Additionally, a current probe (Tektronix TCP2020) was applied across the digital power supply line for prompt measurement of current variations   as a consequence of read-out activity. The absorbed power was directly determined by measuring the current values on the analog and digital supply lines, together with the supply voltage.

The turret was controlled by an engineering model TDAQ, implemented on a Digilent Nexys Video evaluation board, completed with a custom expansion board with all the necessary connectors and interfaces. The TDAQ, in turn, was operated from a laptop PC delivering commands to set up the turret, initialize the ALTAI chips, control the DAQ and perform calibrations.

Trigger pulses were provided to TDAQ by a waveform generator (Rigol DG4202) operated in pulse mode, which allowed to vary the trigger frequency during the test. 
The duration of the clock gate was monitored indirectly by checking the TDAQ output busy signal, which stays asserted for approximately the same time.
The trigger and busy signals were monitored with an oscilloscope (LeCroy HDO 9104-MS), together with the current probe output.

To simulate the passage of a charged particle through the turret, one ALTAI chip per stave was programmed to generate a pattern of nearby hit pixels (known as a \textit{cluster}), by using the internal configurable test-in functionality. The number of pixels composing the cluster (known as cluster size) determines the amount of transferred data and, therefore, the duration of the clock gate and busy signal.



\begin{figure}[ht]
    \centering
    \includegraphics[width = 1 \linewidth]{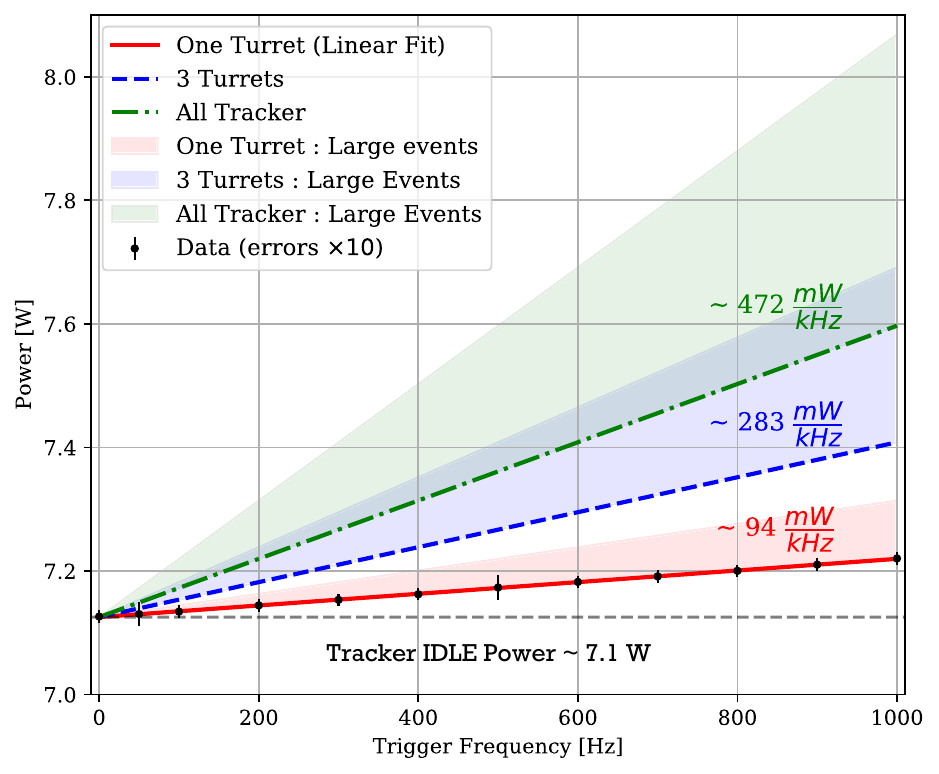}
    \caption{Power consumption of 1, 3 or 5 simultaneously read-out DIR turrets, as a function of trigger frequency. The black points represent the experimental data for DAQ operation of a single turret for small events (cluster size set to 2), superimposed on the baseline DIR power with no triggers. \textcolor{black}{The quoted errors correspond to the standard deviation of repeated measurements.}}
    \label{fig:TrackerPowerConsumptionMeasurement}
\end{figure}

With this experimental setup, we measured the average power consumption of a single turret as a function of trigger frequency up to 1 kHz, for varying cluster sizes. \textcolor{black}{The power measurement was performed through the calibrated voltage and current meters integrated in the power supply unit; with this method, once applied a fixed and continuous trigger rate, the resulting current represents the average value over the periodic bursts of activity initiated by the triggers}. These data were employed to determine the DIR power consumption with DAQ activity on 1, 3 or all 5 turrets, as presented in Fig.~\ref{fig:TrackerPowerConsumptionMeasurement}, \textcolor{black}{by simple multiplication of single turret consumption; the approach was possible since the expected impact of fabrication tolerances on power consumption of involved parts is order of \% (as also verified by measurements) and hence negligible within the scope of the present evaluation.}


The baseline DIR power consumption in idle mode, i.e.~with no triggers at all and therefore no clock gates, was determined to be \textcolor{black}{7.1 W, higher than the 6 W value estimated during design (as reported in Sect. \ref{sect:PowerManagement}), but still acceptable; the difference comes from the approximations of the employed circuital model of ALTAI device, before performing the actual measurement.}

An additional contribution, linearly increasing with trigger frequency, arises from DAQ activity.
As a first step we set the ALTAI cluster size to 2 pixels; this size is typical for the passage of a minimum ionizing particle, such as an electron in the energy range from 3 to 100 MeV. The experimental results for DAQ activity on a single turret are shown in Fig.~\ref{fig:TrackerPowerConsumptionMeasurement} by the black points and the superimposed linear fit. The duration of the busy signal is 70 $\mu$s for most of these events, with some fluctuations up to 100 $\mu$s for a few of them, due to communication latencies related to the firmware version operating in the engineering model TDAQ, as shown in the upper-right panel of Fig.~\ref{fig:PowerCons_Characterization}. The measured increase of power consumption with trigger frequency turns out to be (94 $\pm$ 1) mW/kHz for one read-out turret; this value was extrapolated to the case of 3 and 5 simultaneously read-out turrets, respectively giving (283 $\pm$ 3) mW/kHz and (472 $\pm$ 5) mW/kHz.

Subsequently, we set the ALTAI chips to create larger pixel clusters, typically produced by highly ionizing particles such as nuclei, which would generate longer busy signal durations, up to the maximum allowed (200 µs), saturating the memory space allocated for event data on TDAQ.
 The regions covered by such events are identified with the shaded filled areas in Fig.~\ref{fig:TrackerPowerConsumptionMeasurement}, whose upper limits approximately correspond to twice the slope obtained for minimal cluster sizes.

The maximum DIR power consumption, with all 5 turrets simultaneously read-out, turns out to be less than 8.1 W at 1 kHz trigger frequency. On the other hand, the measured TDAQ power turns out to be 2.6 W at full operation. The combined power thus amounts to 10.7 W, well below the allocated 13 W power budget. \textcolor{black}{This allows to sustain even higher trigger rates (since HEPD-02 can manage peak rates of few kHz) and to take into account in-flight aging of the HEPD-02 power supply unit, which in turn causes loss of efficiency and higher power consumption. In the extreme case of power consumption exceeding the budget, it will be possible to modify the HEPD-02 instrument configuration by suitable telecommands from ground, for example to reduce trigger rates by applying higher pre-scaling factors.}




\section{Conclusion}

The HEPD-02 direction detector is the first space-borne silicon pixel tracker employing the ALTAI Monolithic Active Pixel Sensor (MAPS).

A custom data acquisition (TDAQ) board and firmware have been implemented to exploit the excellent detection performances of the MAPS, while keeping the power consumption within the strict limits imposed by the satellite mission. This original design includes a read-out protocol implemented with the ALTAI slow-control bus, coupled with extensive use of clock gating.

\textcolor{black}{The direction detector and TDAQ board have been successfully integrated in the HEPD-02 flight and qualification models, which underwent extensive tests with particle beams and cosmic rays at ground level.}

Our application shows the feasibility of using MAPS in space with a limited power consumption. This, in turn, opens to the design of large tracking detectors using MAPS, at a scale not previously attainable with strips detectors in terms of costs, complexity and 2-D point resolution \cite{Aladino}. The next generation of MAPS \cite{astropix, arcadia} is currently being designed considering the possibility of space application and consequently implementing built-in power saving solutions.





%

\acknowledgments

This work was supported by the Italian Space Agency (ASI) in the framework of the ``Accordo attuativo 2020-32.HH.0 Limadou Scienza+'' and the ASI-INFN agreement n. 2014-037-R.0, addendum 2014-037-R-1-2017.


\bibliographystyle{JHEP}


\bibliography{biblio}

@article{DeSantis:2021WO,
  author = "De Santis, Cristian  and  Ricciarini, Sergio",
  title = "{The High Energy Particle Detector (HEPD-02) for the second China Seismo-Electromagnetic Satellite (CSES-02)}",
  doi = "10.22323/1.395.0058",
  journal = "PoS",
  year = 2021,
  volume = "ICRC2021",
  pages = "058"
}

@ARTICLE{lentaris,
  author={Lentaris, George and Maragos, Konstantinos and Soudris, Dimitrios and Di Capua, Francesco and Campajola, Luigi and Campajola, Marcello and Costantino, Alessandra and Furano, Gianluca and Tavoularis, Antonios and Santos, Lucana},
  journal={IEEE Transactions on Nuclear Science}, 
  title={{TID} Evaluation System With On-Chip Electron Source and Programmable Sensing Mechanisms on {FPGA}}, 
  year={2019},
  volume={66},
  number={1},
  pages={312-319},
  doi={10.1109/TNS.2018.2885713}}

@article{tambara,
   author={Tambara, Lucas Antunes and Kastensmidt, Fernanda Lima and Medina, Nilberto H. and Added, Nemitala and Aguiar, Vitor A. P. and Aguirre, Fernando and Macchione, Eduardo L. A. and Silveira, Marcilei A. G.},
  journal={2015 IEEE Radiation Effects Data Workshop (REDW)}, 
  title={Heavy Ions Induced Single Event Upsets Testing of the 28 nm {Xilinx Zynq-7000} All Programmable SoC}, 
  year={2015},
  volume={},
  number={},
  pages={1-6},
  doi={10.1109/REDW.2015.7336716}}

@article{Scotti:2019/9,
  author = "Scotti, Valentina  and  Osteria, Giuseppe",
  title = "{The High Energy Particle Detector onboard CSES-02 satellite}",
  doi = "10.22323/1.358.0135",
  journal = "PoS",
  year = 2019,
  volume = "ICRC2019",
  pages = "135"
}

@article{ambrosi2018hepd,
 author = {Ambrosi, Giovanni and Bartocci, Simona and Basara, Laurent and Battiston, Roberto and Burger, William J. and Carfora, Luca and Castellini, Guido and Cipollone, Piero and Conti, Livio and Contin, Andrea and De Donato, Cinzia and De Santis, Cristian and Follega, Francesco M. and Guandalini, Cristina and Ionica, Maria and Iuppa, Roberto and Laurenti, Giuliano and Lazzizzera, Ignazio and Lolli, Mauro and Manea, Christian and Marcelli, Laura and Masciantonio, Giuseppe and Mergé, Matteo and Osteria, Giuseppe and Pacini, Lorenzo and Palma, Francesco and Palmonari, Federico and Panico, Beatrice and Patrizii, Laura and Perfetto, Francesco and Picozza, Piergiorgio and Pozzato, Michele and Puel, Matteo and Rashevskaya, Irina and Ricci, Ester and Ricci, Marc and Ricciarini, Sergio Bruno and Scotti, Valentina and Sotgiu, Alessando and Sparvoli, Roberta and Spataro, Bruno and Vitale, Vincenzo},
	title = {The {HEPD} particle detector of the {CSES} satellite mission for investigating seismo-associated perturbations of the {Van Allen} belts},
	year = {2018},
	journal = {Science China Technological Sciences},
	volume = {61},
	number = {5},
	pages = {643 – 652},
	doi = {10.1007/s11431-018-9234-9},
}

@article{martucci2023first,
 	author = {Martucci, Matteo and Laurenza, Monica and Benella, Simone and Berrilli, Francesco and Del Moro, Dario and Giovannelli, Luca and Parmentier, Alexandra and Piersanti, Mirko and Albrecht, Gabor and Bartocci, Simona and Battiston, Roberto and Burger, William J. and Campana, Donatella and Carfora, Luca and Consolini, Giuseppe and Conti, Livio and Contin, Andrea and De Donato, Cinzia and De Santis, Cristian and Follega, Francesco Maria and Iuppa, Roberto and Lega, Alessandro and Marcelli, Nadir and Masciantonio, Giuseppe and Mergé, Matteo and Mese, Marco and Oliva, Alberto and Osteria, Giuseppe and Palma, Francesco and Panico, Beatrice and Perfetto, Francesco and Picozza, Piergiorgio and Pozzato, Michele and Ricci, Ester and Ricci, Marco and Ricciarini, Sergio Bruno and Sahnoun, Zouleikha and Scotti, Valentina and Sotgiu, Alessandro and Sparvoli, Roberta and Vitale, Vincenzo and Zoffoli, Simona and Zuccon, Paolo},
	title = {The First Ground-Level Enhancement of Solar Cycle 25 as Seen by the High-Energy Particle Detector ({HEPD-01}) on Board the {CSES-01} Satellite},
	year = {2023},
	journal = {Space Weather},
	volume = {21},
	number = {1},
	doi = {10.1029/2022SW003191},
	type = {Article},
}

@article{palma2021high,
  title="{The High-Energy Particle Detector (HEPD-01) as a space weather monitoring instrument on board the CSES-01 satellite}",
  author={Palma, Francesco and Martucci, Matteo and Parmentier, Alexandra and Piersanti, Mirko and Sotgiu, A},
  doi = "10.22323/1.395.1275",
  journal = "PoS",
  year = 2021,
  volume = "ICRC2021",
  pages = "1275"

}

@article{palma2021august,
AUTHOR = { {Palma}, Francesco and Sotgiu, Alessandro and Parmentier, Alexandra and Martucci, Matteo and Piersanti, Mirko and Bartocci, Simona and Battiston, Roberto and Burger, William Jerome and Campana, Donatella and Carfora, Luca and Castellini, Guido and Conti, Livio and Contin, Andrea and D’Angelo, Giulia and De Donato, Cinzia and De Santis, Cristian and Follega, Francesco Maria and Iuppa, Roberto and Lazzizzera, Ignazio and Marcelli, Nadir and Masciantonio, Giuseppe and Mergé, Matteo and Oliva, Alberto and Osteria, Giuseppe and Palmonari, Federico and Panico, Beatrice and Perfetto, Francesco and Picozza, Piergiorgio and Pozzato, Michele and Ricci, Ester and Ricci, Marco and Ricciarini, Sergio Bruno and Sahnoun, Zouleikha and Scotti, Valentina and Sparvoli, Roberta and Vitale, Vincenzo and Zoffoli, Simona and Zuccon, Paolo},
TITLE = {The August 2018 Geomagnetic Storm Observed by the High-Energy Particle Detector on Board the {CSES-01} Satellite},
JOURNAL = {Applied Sciences},
VOLUME = {11},
YEAR = {2021},
NUMBER = {12},
ARTICLE-NUMBER = {5680},
ISSN = {2076-3417},
DOI = {10.3390/app11125680}
}

@article{sgrigna2005correlations,
 author = {Sgrigna, V. and Carota, L. and Conti, L. and Corsi, M. and Galper, A.M. and Koldashov, S.V. and Murashov, A.M. and Picozza, P. and Scrimaglio, R. and Stagni, L.},
	title = {Correlations between earthquakes and anomalous particle bursts from {SAMPEX/PET} satellite observations},
	year = {2005},
	journal = {Journal of Atmospheric and Solar-Terrestrial Physics},
	volume = {67},
	number = {15},
	pages = {1448 – 1462},
	doi = {10.1016/j.jastp.2005.07.008},
}

@ARTICLE{picozza2019hepd,
	author = {Picozza, P. and others},
	title = "{Scientific goals and in-orbit performance of the high-energy particle detector on board the CSES}",
	year = {2019},
	journal = {ApJS},
	volume = {243},
	number = {1},
	doi = {10.3847/1538-4365/ab276c},
	type = {Article},

}

@article{aleksandrin2003high,
 author = {Aleksandrin, S.Yu. and Galper, A.M. and Grishantzeva, L.A. and Koldashov, S.V. and Maslennikov, L.V. and Murashov, A.M. and Picozza, P. and Sgrigna, V. and Voronov, S.A.},
	title = {High-energy charged particle bursts in the near-Earth space as earthquake precursors},
	year = {2003},
	journal = {Annales Geophysicae},
	volume = {21},
	number = {2},
	pages = {597 – 602},
	doi = {10.5194/angeo-21-597-2003},
}

@article{ricciarini2021enabling,
  author = "Ricciarini, Sergio B.  and  Beolé, Stefania  and  de Cilladi, Lorenzo  and  Gebbia, Giuseppe  and  Iuppa, Roberto  and  Ricci, Ester  and  Zuccon, Paolo",
  title = "{Enabling low-power MAPS-based space trackers: a sparsified readout based on smart clock gating for the High Energy Particle Detector HEPD-02}",
  doi = "10.22323/1.395.0071",
  journal = "PoS",
  year = 2021,
  volume = "ICRC2021",
  pages = "071"
}

@article{coli2021sissa,
  author = "Coli, Silvia  and  Angeletti, Massimo  and  Gargiulo, Corrado  and  Iuppa, Roberto  and  Serra, Enrico",
  title = "{Development of a Carbon-fiber reinforced polymer-based mechanics for embedding ALPIDE pixel sensors in the High-Energy Particle Detector space module onboard the CSES-02 satellite.}",
  doi = "10.22323/1.395.0068",
  journal = "PoS",
  year = 2021,
  volume = "ICRC2021",
  pages = "068"
}

@article{iuppa2021innovative,
  author = "Iuppa, Roberto  and  Beolé, Stefania  and  Coli, Silvia  and  Gebbia, Giuseppe  and  Ricci, Ester  and  Ricciarini, Sergio  and  Serra, Enrico  and  Zuccon, Paolo  and  de Cilladi, Lorenzo",
  title = "{The innovative particle tracker for the HEPD space experiment onboard the CSES-02 satellite}",
  doi = "10.22323/1.395.0070",
  journal = "PoS",
  year = 2021,
  volume = "ICRC2021",
  pages = "070"
}

@article{scotti2023trigger,
title = {Trigger and data acquisition system of the High Energy Particle Detector on board the {CSES-02} satellite},
journal = {NIM A},
volume = {1046},
pages = {167741},
year = {2023},
issn = {0168-9002},
doi = {https://doi.org/10.1016/j.nima.2022.167741},
author = {Valentina Scotti and CSES-LIMADOU-Collaboration}
}

@article{masciantonio2021hepd,
  author = "Masciantonio, Giuseppe  and  De Donato, Cinzia  and  Sotgiu, Alessandro",
  title = "{The HEPD-02 Data Processing and Control Unit for the CSES-02 mission}",
  doi = "10.22323/1.395.0059",
  journal = "PoS",
  year = 2021,
  volume = "ICRC2021",
  pages = "059"
}

@article{bates2018single,
	author = {Bates, T. and Bridges, C.P.},
	title = {Single event mitigation for Xilinx 7-series FPGAs},
	year = {2018},
	journal = {IEEE Aerospace Conference Proceedings},
	volume = {2018-March},
	pages = {1 – 12},
	doi = {10.1109/AERO.2018.8396520},
}

@Manual{semxilinx,
  title        = "{LogiCORE IP Soft Error
Mitigation Controller}",
  year         = {2011},
  number       = {DS796},
  note         = {v3.1},
  organization = {Xilinx},
  url          = {https://docs.amd.com/v/u/en-US/ug764_sem},
}

@article{Nicolaidis:2023l9,
  author = "Nicolaidis, Riccardo  and  Gebbia, Giuseppe  and  Iuppa, Roberto  and  Zuccon, Paolo  and  Ricci, Ester  and  Nozzoli, Francesco",
  title = "{The TDAQ system of the HEPD-02 on the CSES-02 mission}",
  doi = "10.22323/1.444.1321",
  journal = "PoS",
  year = 2023,
  volume = "ICRC2023",
  pages = "1321"
}

@ARTICLE{hepd02_dir,
  author={Barioglio, Luca and Bartocci, Simona and Battiston, Roberto and Beolè, Stefania and Benotto, Franco and Bufalino, Stefania and Cipollone, Piero and Coli, Silvia and Contin, Andrea and Cristoforetti, Marco and De Donato, Cinzia and De Santis, Cristian and Luca, Andrea Di and Dumitrache, Floarea and Ferrero, Chiara and Follega, Francesco Maria and Botta, Simone Garrafa and Gebbia, Giuseppe and Iuppa, Roberto and Lega, Alessandro and Lolli, Mauro and Masciantonio, Giuseppe and Mergè, Matteo and Mese, Marco and Nicolaidis, Riccardo and Nozzoli, Francesco and Oliva, Alberto and Osteria, Giuseppe and Palma, Francesco and Palmonari, Federico and Panico, Beatrice and Perciballi, Stefania and Perfetto, Francesco and Picozza, Piergiorgio and Pozzato, Michele and Ricci, Marco and Ricci, Ester and Ricciarini, Sergio Bruno and Sahnoun, Zouleikha and Savino, Umberto and Scotti, Valentina and Serra, Enrico and Sotgiu, Alessandro and Sparvoli, Roberta and Ubertini, Pietro and Vilona, Veronica and Zoffoli, Simona and Zuccon, Paolo},
  journal={IEEE Aerospace and Electronic Systems Magazine}, 
  title={The Monolithic Active Pixel Sensors Tracker System of the High-Energy Particle Detector Aboard the Second Chinese Seismo-Electromagnetic Satellite}, 
  year={2025},
  volume={40},
  number={10},
  pages={28-46},
  doi={10.1109/MAES.2025.3568361}}

@Article{Aladino,
AUTHOR = {Adriani, Oscar and Altomare, Corrado and Ambrosi, Giovanni and Azzarello, Philipp and Barbato, Felicia Carla Tiziana and Battiston, Roberto and Baudouy, Bertrand and Bergmann, Benedikt and Berti, Eugenio and Bertucci, Bruna and Boezio, Mirko and Bonvicini, Valter and Bottai, Sergio and Burian, Petr and Buscemi, Mario and Cadoux, Franck and Calvelli, Valerio and Campana, Donatella and Casaus, Jorge and Contin, Andrea and D’Alessandro, Raffaello and Dam, Magnus and De Mitri, Ivan and de Palma, Francesco and Derome, Laurent and Di Felice, Valeria and Di Giovanni, Adriano and Donnini, Federico and Duranti, Matteo and Fiandrini, Emanuele and Follega, Francesco Maria and Formato, Valerio and Gargano, Fabio and Giovacchini, Francesca and Graziani, Maura and Ionica, Maria and Iuppa, Roberto and Loparco, Francesco and Marín, Jesús and Mariotto, Samuele and Marsella, Giovanni and Martínez, Gustavo and Martínez, Manel and Martucci, Matteo and Masi, Nicolò and Mazziotta, Mario Nicola and Mergé, Matteo and Mori, Nicola and Munini, Riccardo and Musenich, Riccardo and Mussolin, Lorenzo and Nozzoli, Francesco and Oliva, Alberto and Osteria, Giuseppe and Pacini, Lorenzo and Paniccia, Mercedes and Papini, Paolo and Pearce, Mark and Perrina, Chiara and Picozza, Piergiorgio and Pizzolotto, Cecilia and Pospíšil, Stanislav and Pozzato, Michele and Quadrani, Lucio and Ricci, Ester and Rico, Javier and Rossi, Lucio and Schioppa, Enrico Junior and Serini, Davide and Smolyanskiy, Petr and Sotgiu, Alessandro and Sparvoli, Roberta and Surdo, Antonio and Tomassetti, Nicola and Vagelli, Valerio and Velasco, Miguel Ángel and Wu, Xin and Zuccon, Paolo},
TITLE = {Design of an Antimatter Large Acceptance Detector In Orbit ({ALADInO})},
JOURNAL = {Instruments},
VOLUME = {6},
YEAR = {2022},
NUMBER = {2},
ARTICLE-NUMBER = {19},
ISSN = {2410-390X},
doi = {10.3390/instruments6020019}
}

@article{astropix, 
     author = {Striebig, N. and Leys, R. and Peric, I. and Caputo, R. and Steinhebel, A.L. and Suda, Y. and Fukazawa, Y. and Jadhav, M. and Violette, D. and Kierans, C. and Tajima, H. and Metcalfe, J. and Perkins, J.S.},
	title = {AstroPix4 — a novel HV-CMOS sensor developed for space based experiments},
	year = {2024},
	journal = {Journal of Instrumentation},
	volume = {19},
	number = {4},
	doi = {10.1088/1748-0221/19/04/C04010},
	url = {https://www.scopus.com/inward/record.uri?eid=2-s2.0-85189864423&doi=10.1088%2f1748-0221%2f19%2f04%2fC04010&partnerID=40&md5=fcea5da913bfa3f862ebe311561035a9},
	type = {Article},
}

@article{arcadia,
doi = {10.1088/1748-0221/18/02/C02045},
year = {2023},
month = {feb},
publisher = {IOP Publishing},
volume = {18},
number = {02},
pages = {C02045},
author = {Corradino, T. and Dalla Betta, G.F. and Neubüser, C. and Pancheri, L. and on behalf of the ARCADIA collaboration},
title = {ARCADIA MAPS process qualification through the electrical characterization of passive pixel arrays},
journal = {Journal of Instrumentation}
}

@article{ALPAT2010207,
title = {The internal alignment and position resolution of the {AMS-02} silicon tracker determined with cosmic-ray muons},
journal = {{NIM A}},
volume = {613},
number = {2},
pages = {207-217},
year = {2010},
issn = {0168-9002},
doi = {https://doi.org/10.1016/j.nima.2009.11.065},
author = {B. Alpat and G. Ambrosi and Ph. Azzarello and R. Battiston and B. Bertucci and M. Bourquin and W.J. Burger and F. Cadoux and C.F. {da Silva Costa} and V. Choutko and M. Duranti and E. Fiandrini and D. Haas and S. Haino and M. Ionica and R. Ionica and C. Lechanoine-Leluc and M. Menichelli and S. Natale and A. Oliva and M. Paniccia and E. Perrin and M. Pohl and D. Rapin and N. Tomassetti and P. Zuccon and C. Zurbach}
}

@article{STRAULINO2004168,
title = {The {PAMELA} silicon tracker},
journal = {{NIM A}},
volume = {530},
number = {1},
pages = {168-172},
year = {2004},
issn = {0168-9002},
doi = {https://doi.org/10.1016/j.nima.2004.05.067},
author = {S. Straulino and O. Adriani and L. Bonechi and M. Bongi and G. Castellini and R. D'Alessandro and A. Gabbanini and M. Grandi and P. Papini and S.B. Ricciarini and P. Spillantini and F. Taccetti and M. Tesi and E. Vannuccini}
}

@article{ALPIDE_structure,
  title={The {ALPIDE} pixel sensor chip for the upgrade of the {ALICE Inner Tracking System}},
  author={Aglieri Rinella, G and ALICE Collaboration},
  journal={NIM A},
  volume={845},
  pages={583--587},
  year={2017},
  publisher={Elsevier},
  doi={https://doi.org/10.1016/j.nima.2016.05.016}
}

@Article{instruments7040053,
AUTHOR = {Mese, Marco and Anastasio, Antonio and Boiano, Alfonso and Masone, Vincenzo and Osteria, Giuseppe and Perfetto, Francesco and Panico, Beatrice and Scotti, Valentina and Vanzanella, Antonio},
TITLE = {The {PMT} Acquisition and Trigger Generation System of the {HEPD-02} Calorimeter for the {CSES-02} Satellite},
JOURNAL = {Instruments},
VOLUME = {7},
YEAR = {2023},
NUMBER = {4},
ARTICLE-NUMBER = {53},
ISSN = {2410-390X},
DOI = {10.3390/instruments7040053}
}

@misc{ spacewire,
title = {SpaceWire Light},
author = {Ruckman, Larry},
abstractNote = {SpaceWire Light is a SpaceWire encoder-decoder. It is synthesizable for FPGA targets (up to 200 Mbit on Spartan-3). Application interfaces include a simple FIFO interface, as well as an AMBA bus interface for LEON3 system-on-chip designs.},
doi = {10.11578/dc.20220810.1},
url = {https://doi.org/10.11578/dc.20220810.1},
howpublished = {[Computer Software] \url{https://doi.org/10.11578/dc.20220810.1}},
year = {2010},
month = {jun}
}
\end{document}